\documentclass[aps,11pt,superscriptaddress,preprintnumbers,
                secnumarabic,nofootinbib]{revtex4}

\usepackage{epsfig}
\usepackage{bbm}

\begin{document} 

\title{\Large${\rm SO}(10)$-GUT Coherent Baryogenesis}

\author{Bj\"orn~Garbrecht}
\email[]{B.Garbrecht@ThPhys.Uni-Heidelberg.De}
\affiliation{Institut f\"ur Theoretische Physik, Universit\"at Heidelberg, Philosophenweg 16, 69120 Heidelberg, Germany}

\author{Tomislav Prokopec}
\email[]{T.Prokopec@Phys.UU.NL}
\affiliation{Institute for Theoretical Physics (ITF) \& Spinoza Institute, Utrecht University, Leuvenlaan 4, Postbus 80.195, 3508 TD Utrecht, The Netherlands}

\author{Michael G. Schmidt}
\email[]{M.G.Schmidt@ThPhys.Uni-Heidelberg.De}
\affiliation{Institut f\"ur Theoretische Physik, Universit\"at Heidelberg, Philosophenweg 16, 69120 Heidelberg, Germany}

\preprint{HD-THEP-05-21}
\preprint{SPIN-05/26,ITP-UU-05/32}

\begin{abstract}
A model for GUT baryogenesis, coherent baryogenesis within the framework of
supersymmetric ${\rm SO}(10)$, is considered.
In particular, we discuss the Barr-Raby model, where at the end of
hybrid inflation charge asymmetries can be created through the time-dependent
higgsino-gaugino mixing mass matrix. These asymmetries are processed to
Standard Model matter through decays \emph{via} nonrenormalizable
$(B\!-\!L)$-violating operators.
We find that a baryon asymmetry
in accordance with observation can be generated. An appendix is devoted to
provide useful formulas and concrete examples for calculations within
${\rm SO}(10)$.
\end{abstract}


\maketitle

\section{Introduction}

Grand Unified Theories (GUTs) generically predict a scalar potential
and thereby a large amount of vacuum energy when the scalar fields
are displaced from the minimum which breaks the symmetry down to the
Standard Model. It is hence often argued that cosmic inflation
may be implemented by the scalar field dynamics of a GUT~\cite{Guth:1980}.
A well-known paradigm is supersymmetric (SUSY)
hybrid inflation~\cite{DvaliShafiSchaefer:1994,CopelandLiddleLythStewartWands:1994},
and it has led to various models using different grand unified gauge
groups, see \emph{e.g.}~\cite{CoviManganoMasieroMiele:1997,JeannerotKhalilLazaridesShafi:2000,KyaeShafi:2005}.

Another feature of GUTs, which is of possible relevance for cosmology,
is obviously the violation of baryon minus lepton number, $B\!-\!L$,
due to the unification and mixing
of baryons and leptons, since this can lead to mechanisms for
generating the observed baryon asymmetry of the Universe (BAU).
While leptogenesis is strictly speaking not necessarily implemented into a GUT
-- it can be operative within the Standard Model minimally extended by
right handed neutrinos with Majorana mass terms --
we  suggested a scenario relying on baryon-lepton unification,
coherent baryogenesis~\cite{GarbrechtProkopecSchmidt:2004,GarbrechtProkopecSchmidt:2004:2},
and implemented it within a Pati-Salam supersymmetric hybrid inflationary 
model.
We give a brief review of this mechanism and our calculational formalism
inspired by kinetic theory in section~\ref{sec:cbg}.

Being a product group, the Pati-Salam gauge group~\cite{PatiSalam:1974}
gives, strictly speaking, not rise to a GUT, and it is
therefore desirable to devise models based on the proper GUT ${\rm SO}(10)$.
A way of breaking the ${\rm SO}(10)$-symmetry, which is particularly
suitable for
SUSY hybrid inflation, is the mechanism proposed by Barr and Raby.
Kyae and Shafi extended this model by global symmetries which
restrict the superpotential to contain only couplings consistent with
hybrid inflation~\cite{KyaeShafi:2005}. This model 
has however a rather complicated Higgs sector.

The purpose of this paper is to show that coherent baryogenesis
naturally occurs within a SUSY-${\rm SO}(10)$-framework. Therefore,
we want to keep the discussion as simple as possible and use
the minimal superpotential suggested by Barr and Raby, leaving aside
the issues which Kyae and Shafi focus on. In turn, we also point out
that we do not extend the minimal model by \emph{ad hoc} terms just
in order to make our mechanism viable.


In section~\ref{sec:BR} we put together the features of the Barr-Raby
model which are important for our baryogenesis scenario and present
in some detail the derivation of the higgsino-gaugino mixing mass matrix.
We also devote an appendix to the conventions and techniques we apply for
calculations within ${\rm SO}(10)$, with
the intention to make this article self-contained and easily
comprehensible to the reader who is not familiar with ${\rm SO}(10)$-model
building, and furthermore we want to provide a useful help for accessing
the papers mentioned above.

Putting our considerations into work, we present a numerical study
of ${\rm SO}(10)$-coherent baryogenesis in section~\ref{sec:sim}. There,
we also discuss the decay processes of the higgsinos \emph{via}
nonrenormalizable, $(B\!-\!L)$-violating couplings. The result is an
estimate of the produced baryon asymmetry for a particular set of parameters.

\section{Coherent Baryogenesis\label{sec:cbg}}

Coherent baryogenesis relies on the production of particles due
to a time dependent mass term, a phenomenon which we refer to as preheating.
Preheating has been extensively studied for
scalars~\cite{TraschenBrandenberger:1990,KofmanLindeStarobinsky:1997,ProkopecRoos:1996,MichaSchmidt:1999}
and for 
fermions~\cite{GreeneKofman:1998,ChungKolbRiottoTkachev:1999,GiudicePelosoRiottoTkachev:1999},
using the technique of Bogolyubov transformations.
A special
feature of fermionic preheating is that fermions can be amply produced
when their mass term crosses zero.
When one considers instead of a mass term a mixing mass matrix,
charge $C$ and parity $P$ may be violated and an asymmetry can be stored within
different fermionic species. A nonsymmetric mass matrix however leads to
the violation of the orthogonality of particle and antiparticle modes.
One therefore needs a formalism which is independent of a basis in
terms of particle and antiparticle creation and annihilation operators
and thereby generalizes
the Bogolyubov transformation approach. This is developed in
Ref.~\cite{GarbrechtProkopecSchmidt:2002} and shall be briefly reviewed
in the following.

We consider several fermionic flavours, mixing through a mass matrix $M(\eta)$,
which is a function of the conformal time $\eta$.
In a spatially flat Friedmann-Lema\^itre-Robertson-Walker Universe,
described by the metric
$g_{\mu\nu}=a^2(\eta)\times{\rm diag}(1,-1,-1,-1)$,
we rescale the fields such that for the mass terms, there is the
replacement $M(\eta) \rightarrow a(\eta) M(\eta)$.

Our goal is to compute the charge density, which is a bilinear form in
the fermionic fields. We therefore introduce the Wigner function
\begin{eqnarray}
\label{Wigner:MultiFlavour}
   {\rm i}S^<_{ij}(k,x)
 = -\int d^4r {\rm e}^{{\rm i}k\cdot r}
   \langle \bar{\psi}_{j}(x-r/2)\psi_{i}(x+r/2) \rangle
,
\end{eqnarray}
where $i$, $j$ are flavour indices and
$({\rm i}\gamma^0S^{<})^\dagger = {\rm i}\gamma^0S^{<}$ is hermitean.
The Wigner transform is the Fourier transformation of the two point
function \emph{w.r.t.} its relative coordinate while keeping the center
of mass coordinate fixed. One can hence consider it as an analogue to a
classical phase space density defined in quantum theory. Since we
assume here
spatial homogeneous conditions, one can ignore the center of mass coordinate
in the following and consider  ${\rm i}S^<$ as a Fourier transform.
Our formalism is applicable for a 2-point
function with general density matrix, but in view of our applications in
inflation we prefer to write it with respect to the vacuum from the outset.

When decomposing the mass matrix $M$ into
its hermitean and antihermitean parts,
\noindent\vskip -0.15in
\begin{eqnarray}
M_H=\frac{1}{2}\left(M+M^{\dagger}\right),\qquad
M_A=\frac{1}{2 {\rm i}}\left(M-M^{\dagger}\right)
\,,
\nonumber
\end{eqnarray}
\vskip -0.05in\noindent
we find that ${\rm i}S^{<}$ obeys the Wigner space Dirac equation
\begin{equation}
\Bigl( {k}\!\!\!/\; + \frac{\rm i}{2}\gamma^0 \partial_{\eta}
    - (M_{H}+{\rm i} \gamma^5 M_{A})
      {\rm e}^{-\frac{\rm i}{2}\stackrel{\!\!\leftarrow}{\partial_{\eta}}\partial_{k_0}}
\Bigr)_{\!\!il} {\rm i}S_{lj}^{<} = 0
\label{DiracEq}
\,.
\end{equation}

The mass matrix $M$ emerges generically from Yukawa couplings to scalar field
condensates, ${\cal L}_{\rm Yu} = -y\phi\bar\psi_R\psi_L + {\rm h.c.}$ In the
model we consider here, $M$ is the higgsino-gaugino mixing mass matrix.

A crucial point is the time-dependence of $M$, which is not only the
source of particle production.
The matrices $M$ and  ${d}M/{d\eta}$ both contribute to
$CP$-violating phases, which -- provided $M$ and ${d}M/{d\eta}$ 
are linearly independent -- can not be removed 
by time-independent redefinitions of the fermionic fields.

In order to simplify the Wigner-Dirac equation~(\ref{DiracEq}),
which is, besides the flavour indices, also endowed with a
$4\times4$ spinor structure,
we make use of the fact that for spatially homogeneous 
$i\gamma_{0}S^{<}_{h}$, the helicity operator
$
\hat{h} = \hat{\mathbf{k}}\cdot\gamma^{0}
          \mbox{\boldmath{{$\gamma$}}}\gamma^{5}
$
commutes with the Dirac operator in (\ref{DiracEq}) and decompose the Wigner
function as~\cite{KainulainenProkopecSchmidtWeinstock:2001,ProkopecSchmidtWeinstock:2003}
\begin{equation}
-i\gamma_{0}S^{<}_{h}
  = \frac{1}{4}\bigl(\mathbbm{1}+h\hat{\mathbf{k}}\cdot
                     \mbox{\boldmath{{$\sigma$}}}
               \bigr)\otimes\rho^{\mu}g_{\mu h}
,
\label{S<:helicity-diagonal}
\end{equation}
where we have omitted the flavour indices,  
$\hat{\mathbf{k}} = \mathbf{k}/|\mathbf{k}|$
and $\sigma^{\mu}$, $\rho^{\mu}$ ($\mu=0,1,2,3$) are the Pauli matrices,
and $h=\pm 1$ are the eigenvalues of $\hat h$. 
We multiply~(\ref{DiracEq}) by $\rho^{\mu}$, take the Dirac trace and integrate
the hermitean part over $k_0$.
Introducing the 0th momenta of $g_{\mu h}$,
 $f_{\mu h}=\int({dk_0}/{2\pi})g_{\mu h}$,
we note that the functions $f^{ij}_{\mu h}$ explicitly read
\begin{eqnarray}
f^{ij}_{0h}(x,\mathbf{k}) \!\!\!&=&\!\!\! - \int \frac{dk_0}{2\pi}\int d^4r\, {\rm e}^{{\rm i}k\cdot r}
   \langle \bar{\psi}_{hj}(x-r/2) \gamma^0 \psi_{hi}(x+r/2)\rangle\,,
\\
f^{ij}_{1h}(x,\mathbf{k}) \!\!\!&=&\!\!\! -\int \frac{dk_0}{2\pi}\int d^4r\, {\rm e}^{{\rm i}k\cdot r}
   \langle \bar{\psi}_{hj}(x-r/2)  \psi_{hi}(x+r/2)\rangle\,,
\nonumber\\
f^{ij}_{2h}(x,\mathbf{k}) \!\!\!&=&\!\!\! -\int \frac{dk_0}{2\pi}\int d^4r\, {\rm e}^{{\rm i}k\cdot r}
   \langle \bar{\psi}_{hj}(x-r/2) (-{\rm i}\gamma^5)\psi_{hi}(x+r/2)\rangle\,,
\nonumber\\
f^{ij}_{3h}(x,\mathbf{k}) \!\!\!&=&\!\!\! -\int \frac{dk_0}{2\pi}\int d^4r\, {\rm e}^{{\rm i}k\cdot r}
   \langle 
           \bar{\psi}_{hj}(x-r/2) \gamma^0\gamma^5 \psi_{hi}(x+r/2)
   \rangle\,.
\nonumber
\end{eqnarray}
Therefore, the $f_{\mu h}(x,\mathbf{k})$ can be interpreted as follows:
$f_{0h}$ is the charge density, $f_{3h}$ 
is the axial charge density, and $f_{1h}$ and $f_{2h}$ correspond
to the scalar and pseudoscalar density, respectively.

From the Wigner-Dirac equation~(\ref{DiracEq}), we can now
derive the following system of equations:
\begin{eqnarray}
{f}^\prime_{0h} + {\rm i}\left[M_H,f_{1h}\right] +{\rm i}\left[M_A,f_{2h}\right]
\!\!\!&=&\!\!\! 0\,,
\label{f0eqN}
\\
{f}^\prime_{1h} + 2h|\mathbf{k}|f_{2h} + {\rm i}\left[M_H,f_{0h}\right] -\left\{M_A,f_{3h}\right\}
\!\!\!&=&\!\!\! 0\,,
\label{f1eqN}
\nonumber\\
{f}^\prime_{2h} - 2h|\mathbf{k}|f_{1h}  + \left\{M_H,f_{3h}\right\} +{\rm i}\left[M_A,f_{0h}\right] 
\!\!\!&=&\!\!\! 0\,,
\label{f2eqN}
\nonumber\\
{f}^\prime_{3h} - \left\{M_H,f_{2h}\right\} +\left\{M_A,f_{1h}\right\}
\!\!\!&=&\!\!\! 0
\nonumber
\,,
\end{eqnarray}
where the prime denotes a derivative with respect to $\eta$.
It is understood that $M$ and the $f_{\mu h}$ are flavour matrices.
Note that the commutators in~(\ref{f0eqN}), which mix particle flavours, 
are essential for the production of the charges $f_{0h}$, and
thus for our scenario. Moreover, one can infer, that a necessary condition
for $f_{0h}^\prime \not= 0$ is a nonsymmetric $M$. We already anticipated
this when noting that for such a mass matrix the orthogonality of particle
and antiparticle modes is violated.
 Note that the tree level dynamics given by Eqns.~(\ref{f0eqN}) 
closes for $f_{\mu h}$. When rescatterings, as described 
through nonlocal quantum loop corrections, are included, 
off-shell effects may become important, and one would have 
to solve for the full dynamics of the $g_{\mu h}$.
In the nonrelativistic regime
and close to equilibrium however, in which off-shell effects are suppressed,
it is possible to include rescatterings into Eqns.~(\ref{f0eqN}) 
and still retain closure for the equations for the $f_{\mu h}$.

Now we fix the initial conditions for a Universe, which is void of
fermions at the end of inflation.
For an initially diagonal slowly evolving mass matrix,
the Wigner functions for a zero particle state
zero particles are (\emph{cf.} Ref.~\cite{GarbrechtProkopecSchmidt:2002}):
\begin{eqnarray}
 f^{ab}_{0h} &=& L_{h}^{a*}L_{h}^b+R_{h}^{a*}R_h^b,
\quad 
 f^{ab}_{1h} = -2\Re(L_{h}^aR_{h}^*ab),
\nonumber\\
 f^{ab}_{3h} &=& L_{h}^{a*}L_{h}^b-R_{h}^{a*}R_h^b,
\quad
 f^{ab}_{2h} = 2\Im(L_{h}^{*a}R_{h}^{b})
,
\label{f3asLR}
\end{eqnarray}
with
\begin{eqnarray}
L_h^{ab}=\delta_{ab}\sqrt{\frac{\omega_{a}+hk}{2\omega_{a}}},\quad
R_h^{ab}=\delta_{ab}\frac{M_{aa}^*}{\sqrt{2\omega_{a}(\omega_{a}+hk)}}
,
\nonumber
\end{eqnarray}
where $\omega_{a}=\sqrt{\mathbf{k}^2+|M_{aa}|^2}$. Note, that for the
case of real $M_{aa}$, this reduces just to the usual choice of
the components of the basis spinors in chiral representation.
For a nondiagonal, but hermitean, $M$, one obtains the initial conditions by an
appropriate unitary transformation. If additionally $M_A\not=0$, as is
the case for the ${\rm SO}(10)$ example discussed in the following, a biunitary
transformation is necessary for diagonalization.

Since $f_{0h}$ is the zeroth component
of the vector current, the charge of  the species $a$ carried
by the mode with momentum $k$ and helicity $h$
is simply $q_{ah}(k)\!=\!f_{0h}^{aa} -1 $.
Note also, that the Lagrangean
\noindent\vskip -0.2in
\begin{eqnarray}
{\cal L}=
\bar{\psi}_{a}\partial\!\!\!/\, \psi_{a} - \bar{\psi}_{b} (M_H + i \gamma^5 M_A)_{ba} \psi_{a}
\nonumber
\end{eqnarray}\noindent
is $U(1)$ symmetric, and thus $\sum_a q_{ah}(k)$ is conserved, as
we shall verify explicitly for the ${\rm SO}(10)$ example discussed here.

The scenario for coherent baryogenesis is as follows:
initially, there are zero fermions described by appropriate initial
conditions for the $f_{\mu h}$, and $M$ is approximately constant in time.
Then a phase transition occurs during which $M$ changes rapidly, which
leads to fermion production. Eventually, $M$ stops evolving and the
produced number of fermions as well as the charges $f_{0h}$
stored within the different species are frozen in. We emphasise that 
$f_{0h}^{ii}$ should not be confused with the number of 
produced particle pairs at preheating, 
which in our language can be expressed in terms of the
$f_{ih}$ ($i=1,2,3$) as given in Ref.~\cite{GarbrechtProkopecSchmidt:2002}.

\section{The Barr-Raby Model\label{sec:BR}}

One possibility to break $\textnormal{SO}(10)$ down to the
Standard Model is to use a Higgs multiplet $A$ in the adjoint representation
${\bf 45}$ and another pair of Higgses $C$ and $\bar C$ 
in the spinor representations ${\bf 16}$ and ${\bf \overline{16}}$.
The apparently most simple implementation of this pattern of symmetry breaking,
which is in accordance with particle physics observations,
has been suggested by Barr and Raby~\cite{BarrRaby:1997}.
In the following, we review the features of this model as far as they are
relevant for our baryogenesis scenario, in particular we present
in some detail the derivation of the higgsino-gaugino mass matrix. In
the appendix, we give account of the conventions
we use, in particular how the charges under the Standard Model group
\begin{equation}
G_{SM}={\rm SU}(3)_C\times{\rm SU}(2)_L\times{\rm U}(1)_Y
\label{GSM}
\end{equation}
are assigned to the various multiplets of ${\rm SO}(10)$. 

We consider the  superpotential
\begin{eqnarray}
\label{superpotential:BarrRaby}
W \!\!\!&\supset&\!\!\! \kappa S (C\bar{C}-\mu^2) 
+\frac{\alpha}{4 M_S} {\rm tr} A^4 + \frac 12 M_A {\rm tr} A^2
+ T_1 A T_2 + M_T T_2^2\\
&&+ \bar{C}^\prime[\zeta\frac{PA}{M_S}+\zeta_Z Z_1]C
+ \bar{C}[\xi\frac{PA}{M_S}+\xi_Z Z_2]C^\prime
+ M_{C^\prime} C^\prime\bar{C}^\prime
\,,\nonumber
\end{eqnarray}
where the additional fields $S,P,Z_1,Z_2$ are singlets, $T_1$ and $T_2$
${\bf 10}$-plets of $\textnormal{SO}(10)$. Furthermore, there are the
spinor $C^\prime$ and the conjugate spinor $\bar{C}^\prime$.

Let us first discuss the purely adjoint sector.
The potential is at its minimum, when the condition
\begin{equation}
-F_A^*=\frac{\partial W}{\partial A} = 0
\end{equation}
is met. When $\langle A \rangle={\rm diag}(a_1,a_2,a_3,a_4,a_5)
\otimes {\rm i}\sigma_2$, it follows
\begin{equation}
\label{FTerm:A}
\frac{\alpha}{M_S} a_i^3 - M_A a_i =0\,.
\end{equation}
This can be solved by either $a_i=0$, or $a_i=a$, where
\begin{equation}
\label{a:VEV}
a=\pm\sqrt{\frac{M_A M_S}{\alpha}}.
\end{equation}
In order to step towards the Standard Model, it is possible to
break $\textnormal{SO}(10)$ down to the left-right symmetric group
\begin{equation}
G_{LR}=
\textnormal{SU}(3)_C\times
\textnormal{SU}(2)_L\times
\textnormal{SU}(2)_R\times
\textnormal{U}(1)_{B-L}
\end{equation}
by choosing for $\langle A \rangle$ the Dimopoulos-Wilczek (DW) form
\begin{equation}
\label{DW:form}
\langle A \rangle =
\left(
\begin{array}{ccccc}
a&&&&\\
&a&&&\\
&&a&&\\
&&&0&\\
&&&&0
\end{array}
\right)
\otimes {\rm i}\sigma_2\,.
\end{equation}
Note, that $\langle A \rangle$ being of DW form is proportional to
the $(B-L)$ operator given in Eqn.~(\ref{charge:operators}). In the
appendix, we give account of the explicit construction of the tensor- and
spinor representations of ${\rm SO}(10)$ and the conventions we use.

The two Higgs doublets of the MSSM are contained within $T_1$ and
are identified with the four components which remain massless
by the superpotential~(\ref{superpotential:BarrRaby}) when using the
DW-form for $\langle A \rangle$.
The additional six degrees of freedom of $T_1$, two colour triplets,
become heavy and hence invisible at low energies, such that there
is doublet-triplet splitting.
The second ${\bf 10}$-plet
$T_2$ becomes necessary since a direct mass-term for the triplet components
of $T_1$ would lead to a disastrous rapid higgsino-mediated proton decay.

The Higgs fields $C$ and $\bar C$ reduce the $\textnormal{SO}(10)$ symmetry to
$\textnormal{SU}(5)$. When minimizing the scalar potential,
the absolute values of their VEVs are
$\langle C\rangle = \langle \bar C \rangle =\mu$,
and they point in the
$\textnormal{SU}(5)$-singlet direction with the quantum numbers of a
right-handed neutrino.

Both sectors, the spinorial and the adjoint, in combination reduce
the $\textnormal{SO}(10)$-symmetry to the Standard Model group $G_{SM}$.
However, they need
to be linked together in order to get a congruency of the assignment
of Standard Model quantum numbers and to remove all pseudo-Goldstone
modes from the particle spectrum. The obvious candidate term to add
to the superpotential, $\bar{C} A C$, however destabilizes the
DW form~(\ref{DW:form}) by altering the expression for the
$F$-term~(\ref{FTerm:A})
when the spinors get a nonzero VEV. Barr and Raby
therefore suggested to add the additional spinors $C'$ and $\bar C'$,
which get a zero VEV. The conditions for potential minimization now become
\begin{eqnarray}
-F^*_{\bar{C}^\prime}=
\left[
\zeta \frac{PA}{M_S} + \zeta_Z Z_1
\right]
C + M_{C'}C' =0\,,\\
-F^*_{C^\prime}=
\bar{C}
\left[
\xi \frac{PA}{M_S} + \xi_Z Z_2
\right] + M_{C'}\bar C' =0\,.
\end{eqnarray}

When comparing with Eqn.~(\ref{charge:operators}), we note that
in the DW-form~(\ref{DW:form}) we can identify
$\langle A \rangle\equiv \frac 32 {\rm i}  a (B-L)$.
If we assume that the VEV of $P$ is fixed, then $Z_1$ and $Z_2$ settle to
\begin{eqnarray}
\label{Z_1:VEV}
Z_1=-\frac 32 {\rm i} \zeta/\zeta_Z \frac{\langle P \rangle a}{M_S}\,,\\
\label{Z_2:VEV}
Z_2=-\frac 32 {\rm i} \xi/\xi_Z \frac{\langle P \rangle a}{M_S}\,,
\end{eqnarray}
since $C$ and $\bar C$ point into the right-handed neutrino 
direction, where $B-L=1$.
We have hence achieved a link between the spinorial and adjoint sector
without changing the form of $-F_A^*$.


In our model, $CP$-violation arises from the phase between $\zeta$
and $\xi$ and therefore from couplings of the adjoint to the spinor multiplets.
Let us label the multiplets of the Standard Model group~(\ref{GSM}) by $K$.
The representations
${\bf 16}$ and ${\bf 45}$
harbour as multiplets with common $G_{SM}$ quantum numbers
$K=({\bf 3},{\bf 2},\frac 16)$, $K=({\bf \bar 3},{\bf 1},-\frac{2}{3})$ and
$K=({\bf 1},{\bf 1},1)$.
These multiplets therefore mix through the higgsino mass matrix.
The corresponding conjugate multiplets in
${\bf \overline{16}}$ and ${\bf 45}$ are labeled by $\bar K$. Furthermore,
all these representations contain the singlet $({\bf 1},{\bf 1},0)$.

The spinor pair with 32 complex degrees of freedom breaks the 45-dimensional
$\textnormal{SO}(10)$ down to the 24-dimensional $\textnormal{SU}(5)$. The
21 Goldstone modes come from the multiplets
$K=({\bf 3},{\bf 2},\frac 16)$, $K=({\bf \bar 3},{\bf 1},-\frac{2}{3})$,
$K=({\bf 1},{\bf 1},1)$ plus one linear combination of the singlets
$K=({\bf 1},{\bf 1},0)$ within
${\bf 16}$ and ${\bf \overline{16}}$. The 45-dimensional adjoint reduces the
$\textnormal{SO}(10)$-symmetry to the 15-dimensional $G_{LR}$. Because
of the DW VEV being proportional to the $(B-L)$ operator, the 30
Goldstone modes can be identified with the multiplets
for which $B-L\not= 0$, that are all colour triplets.

Hence, by the supersymmetric Higgs-mechanism, there is a mixing of the
higgsino modes with the gaugino sector, through the Lagrangean terms
\begin{equation}
\label{higgsino-gaugino}
\sqrt{2}g \varphi^* T^a \psi \lambda^a + {\rm h.c.}\,,
\end{equation}
where
$T^a$ is a generator of $\textnormal{SO}(10)$,
normalized as ${\rm tr}\,{(T^a)}^2=1$, $\lambda^a$ a gaugino
and $\varphi$ the scalar superpartner of the $\psi$-fermion.
This will induce higgsino-gaugino mixing mass terms
for both multiplets, $A$ and $C$ when
$K=({\bf 3},{\bf 2},\frac 16)$ or $K=({\bf \bar 3},{\bf 1},-\frac{2}{3})$,
and only for $C$, when $K=({\bf 1},{\bf 1},1)$. 

Let us consider possible mass terms involving only the adjoint Higgs.
If we denote the components of either $({\bf 3},{\bf 2},\frac 16)$ or
$({\bf \bar 3},{\bf 1},-\frac{2}{3})$ by $b_K$, we have
(\emph{cf.} appendix~\ref{app:SO10})
\begin{eqnarray}
{\rm tr}\,A^2\!\!\!&=&\!\!\!-6a^2-2b_K b_{\bar K}\,,\\
{\rm tr}\,A^4\!\!\!&=&\!\!\!6 a^4+4 a^2 b_K b_{\bar K} + b_K^2 b_{\bar K}^2\,.
\end{eqnarray}
Hence, the portion
$\frac{\alpha}{4 M_S} {\rm tr} A^4 + \frac 12 M_A {\rm tr} A^2$
of the superpotential~(\ref{superpotential:BarrRaby}) gives for these modes
a zero mass term
\begin{equation}
m_K=\frac{\alpha a^2}{M_S}-M_A=0\qquad
\textnormal{for}\,\,
K=({\bf 3},{\bf 2},\frac 16)\,\,\textnormal{and}\,\,
K=({\bf \bar 3},{\bf 1},-\frac{2}{3})\,,
\end{equation}
where we have used the VEV~(\ref{a:VEV}) for $a$. This result is expected,
since the multiplets in question are Goldstone. In contrast, we find
\begin{eqnarray}
m_K\!\!\!&=&\!\!\!\frac{\alpha a^2}{M_S}\qquad\textnormal{for}\,\,
K=({\bf 1},{\bf 1},1)\,,({\bf 1},{\bf 3},0)\,,\\
m_K\!\!\!&=&\!\!\!-2 \frac{\alpha a^2}{M_S}\qquad\textnormal{for}\,\,
K=({\bf 8},{\bf 1},0)\,.
\end{eqnarray}

Let us now discuss the mixing of the adjoints and spinors.
$\psi_{A_K}$ and $\psi_{C^\prime_{\bar K}}$ get a mixing mass term through
\begin{equation}
\label{W_A,C^prime}
\frac{\delta^2 W}{\delta A_{\bar K} \delta C^\prime_{\bar K}} =
\xi  \frac{\langle \bar C\rangle \langle P \rangle}{\sqrt 2 M_S}\,.
\end{equation}
The derivation of this term is instructive and works as follows:
In the block-diagonal basis, we have
\begin{equation}
A^{\textnormal{\tiny BL}}_\mathbf{\overline{10}}=\left(
\begin{array}{cc}
0 & 0 \\
\mathbf{\overline{10}} & 0
\end{array}\right)\,,
\end{equation}
which transforms according to~(\ref{blocktrafo}) to the off-diagonal
basis as
\begin{equation}
A_\mathbf{\overline{10}}=
U_{\textnormal{\tiny BLOCK}}^{-1}
A^{\textnormal{\tiny BL}}_\mathbf{\overline{10}}
U_{\textnormal{\tiny BLOCK}}=
\frac 12
\left(
\begin{array}{cc}
\mathbf{\overline{10}} & {\rm i}\mathbf{\overline{10}} \\
{\rm i}\mathbf{\overline{10}} & - \mathbf{\overline{10}}
\end{array}\right)\,.
\end{equation}
The single degrees of freedom $\mathbf{\overline{10}}$ are represented by
the ten antisymmetric $5\times5$ matrices with two nonvanishing entries
of the value $1/\sqrt 2$.
Without loss of generality,
we pick the matrix with $-1/\sqrt 2$ in the first row, fourth column,
corresponding to one degree of freedom of the
$\bar{K}=(\mathbf{\bar{3}},\mathbf{2},-\frac16)$-multiplet.
In order to let the tensor $A$ act on a spinor, we make use of
Eqn.~(\ref{Spinor:Tensor}) and  represent it in terms of
the $\Gamma$-operators as
\begin{equation}
A=-\frac{\sqrt 2}{16}\left(
\left[\Gamma_1,\Gamma_4\right]
+{\rm i}\left[\Gamma_1,\Gamma_9\right]
+{\rm i}\left[\Gamma_6,\Gamma_4\right]
-\left[\Gamma_6,\Gamma_9\right]
\right).
\end{equation}
When paired with the spinor
\begin{equation}
\Psi=\frac 12 \left[
\chi_2^\dagger \chi_3^\dagger \chi_5^\dagger
- \chi_3^\dagger \chi_2^\dagger \chi_5^\dagger
\right]|0\rangle,
\end{equation}
a $G_{SM}$ singlet is formed, and after anticommuting the $\chi_i$-operators,
we find
\begin{equation}
\langle 0 | \chi_1 \chi_2 \chi_3 \chi_4 \chi_5 A \Psi = \frac1{\sqrt 2}\,,
\end{equation}
from which we immediately obtain Eqn.~(\ref{W_A,C^prime}).
The higgsino-gaugino mixing terms can be derived from~(\ref{higgsino-gaugino})
in a very similar way.

Finally, for the mixing among the spinors we have
\begin{equation}
\frac{\delta^2 W}{\delta {\bar C}_{\bar K} \delta C^\prime_K} =
\xi  \alpha_K \frac{a \langle P \rangle}{M_S}\,,
\end{equation}
where $\alpha_K=\frac 32 {\rm i} \left[(B-L)_K-1\right]$ or explicitly,
\begin{equation}
\alpha_K=\left\{
\begin{array}{ccc}
-1 & \textnormal{for} & K=({\bf 3},{\bf 2},\frac 16)\\
-2 & \textnormal{for} & K=({\bf \bar 3},{\bf 1},-\frac 23)\\
0 & \textnormal{for} & K=({\bf 1},{\bf 1},1)
\end{array}
\right.
\,.
\end{equation}
We have used here the VEVs~(\ref{Z_1:VEV},~\ref{Z_2:VEV}) and again the
proportionality of $\langle A \rangle$ to the $(B-L)$ operator.

We are now in the position to write down the higgsino-gaugino mass matrix:
\begin{eqnarray}
\label{SO10:mass:matrix}
&&
\left(
\begin{array}{cccc}
\psi_{\lambda_K} & \psi_{A_K} & \psi_{C_K} & \psi_{C^\prime_K}
\end{array}
\right)\\
&&
\times
\left(
\begin{array}{cccc}
0 & - {\rm i}\sqrt 2 \gamma_K g a &  g \langle C\rangle & 0 \\
{\rm i}\sqrt 2 \gamma_K g a & m_K & 0 &
\xi \frac{\langle \bar C\rangle \langle P \rangle}{\sqrt 2 M_S}\\
g \langle \bar C \rangle & 0 & \kappa \langle S \rangle
&{\rm i}\alpha_K \xi \frac{a\langle P\rangle}{M_S}\\
0 & \zeta \frac{\langle \bar C\rangle \langle P \rangle}{\sqrt 2 M_S}
&{\rm i}\alpha_K \zeta \frac{a\langle P\rangle}{M_S} &M_{C^\prime}\\
\end{array}
\right)
\left(
\begin{array}{c}
\psi_{\lambda_{\bar{K}}}
\\ \psi_{A_{\bar{K}}}
\\ \psi_{{\bar C}_{\bar{K}}}\\
\psi_{{\bar C}^\prime_{\bar{K}}}
\end{array}
\right)
+{\rm h.c.}
\nonumber
\,,
\end{eqnarray}
where
\begin{equation}
\gamma_K=\left\{
\begin{array}{ccc}
\frac 12 & \textnormal{for} & K=({\bf 3},{\bf 2},\frac 16)\\
1 & \textnormal{for} & K=({\bf \bar 3},{\bf 1},-\frac 23)\\
0 & \textnormal{for} & K=({\bf 1},{\bf 1},1)
\end{array}
\right.
\,.
\end{equation}
The mass matrix is nonsymmetric, therefore being endowed
with the necessary prerequisites for coherent baryogenesis.

\section{Simulation of Coherent Baryogenesis\label{sec:sim}}

The superpotential~(\ref{superpotential:BarrRaby})
is of the type suitable for hybrid inflation. We assume that
symmetry breaking by the adjoint sector has already taken place
before or during inflation and is preserved throughout the subsequent history
of the Universe, such that possible monopoles are diluted.
We therefore do not consider the dynamics of the field $A$.
For a discussion of the role of cosmic strings formed at the transition 
$G_{LR}\rightarrow G_{SM}$ 
after inflation, we refer to Ref.~\cite{JeannerotPostma:2005}.

The part of the scalar potential relevant for hybrid inflation reads
\begin{equation}
V=\kappa^2 \left| C^2 - \mu^2\right|^2 + 2 \kappa^2 |S C|^2
\,,
\end{equation}
where we have used $C=\bar C^*$ due to the vanishing of the $D$-terms and
have written $C\equiv \bar C$.

\begin{figure}[htbp]
\begin{center}
\epsfig{file=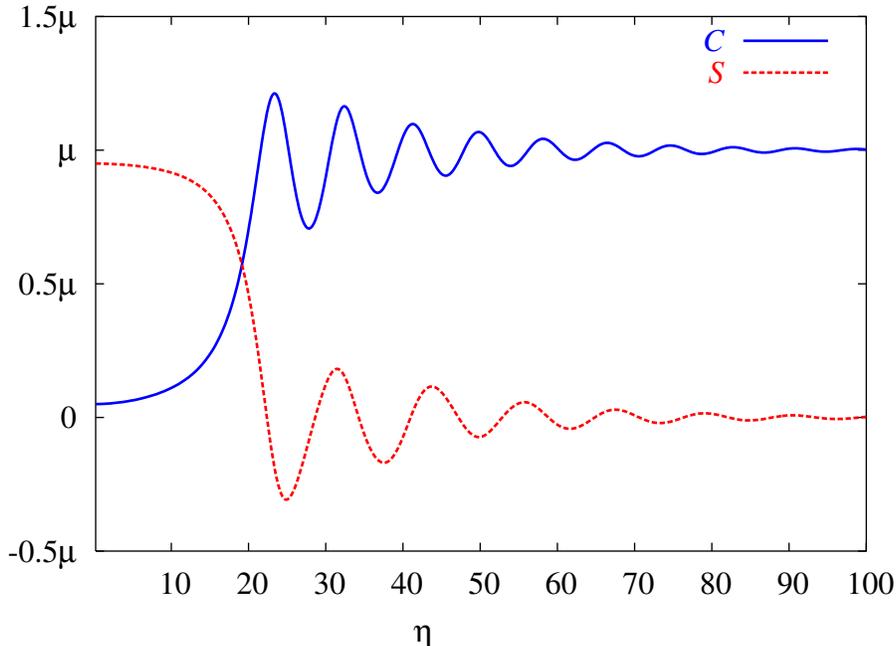, height=3.5in}
\end{center}
\vskip -0.35in
\caption[fig0]{
\label{figure:SO10:ScalMot}
\small
Epoch of phase transition in the $\textnormal{SO}(10)$-model
}
\end{figure}

During inflation, the VEVs of $C$ and $\bar C$ are
sitting at a minimum located at \emph{zero}, and $S$ rolls down
a logarithmic slope until reaching the critical value
$
S_{cr}= \mu
$,
such that the value \emph{zero} for $C$ and $\bar C$ becomes a maximum.
The waterfall regime begins, at the end of which the scalar fields settle
down to the supersymmetric ($V=0$) minimum $S=0$,
$|C|=|\bar C|=\mu$. This is a rapid phase transition which
brings coherent baryogenesis along.
We simulate this scenario for the parameter $\kappa=0.05$, a damping rate
$\Gamma=0.02\mu$
and also take account of the expansion of the Universe,
\emph{cf.} FIG.~{\ref{figure:SO10:ScalMot}}.

Damping partly comes into play because of the perturbative decay of the
inflaton. More important at the beginning of the waterfall regime is
however the phenomenon of tachyonic preheating: Since the fields $C$ and
$\bar C$ attain a negative mass square term, modes with momenta less than
this mass get produced exponentially fast. Here we mimic this effect
by introducing the damping rate $\Gamma$.
For numerical studies of this
process, see Refs.~\cite{GreeneProkopecRoos:1997,Prokopec:1997,FelderEtAl:2000,Garcia-BellidoRuizMorales:2001,CopelandPascoliRajantie:2002,PodolskyFelderKofmanPeloso2005,MichaTkachev:2004}. Damping will also receive a contribution from
fermionic preheating. A proper treatment of fermionic preheating,
which includes rescatterings, would require techniques used in
Ref.~\cite{BergesBorsanyiSerreau:2002}, which have so far not been 
applied to the question of inflaton thermalisation through decay into 
fermions.

In order to keep the discussion simple, we do not take the dynamics of the
singlet fields $Z_1$ and $Z_2$ into account here. In principle, their
VEVs only get fixed when $C$ and
$\bar C$ acquire nonzero VEVs. A possible way to fix  $Z_1$ and $Z_2$ already
during inflation is for example to shift the spinors away from the zero VEV,
as proposed in Ref.~\cite{JeannerotKhalilLazaridesShafi:2000}
and is also applicable to ${\rm SO}(10)$-models~\cite{KyaeShafi:2005}.


For the remaining parameters, we choose
$\mu=0.5\times 10^{16}{\rm GeV}$, $M_S=550\mu$, $M_{C^\prime}=0.02\mu$,
$g=0.2$,
$\zeta=-0.02$, $\xi=0.05{\rm i}$, $a=25\mu$ and $\langle P \rangle=24\mu$.

\begin{figure}[htbp]
\begin{center}
\epsfig{file=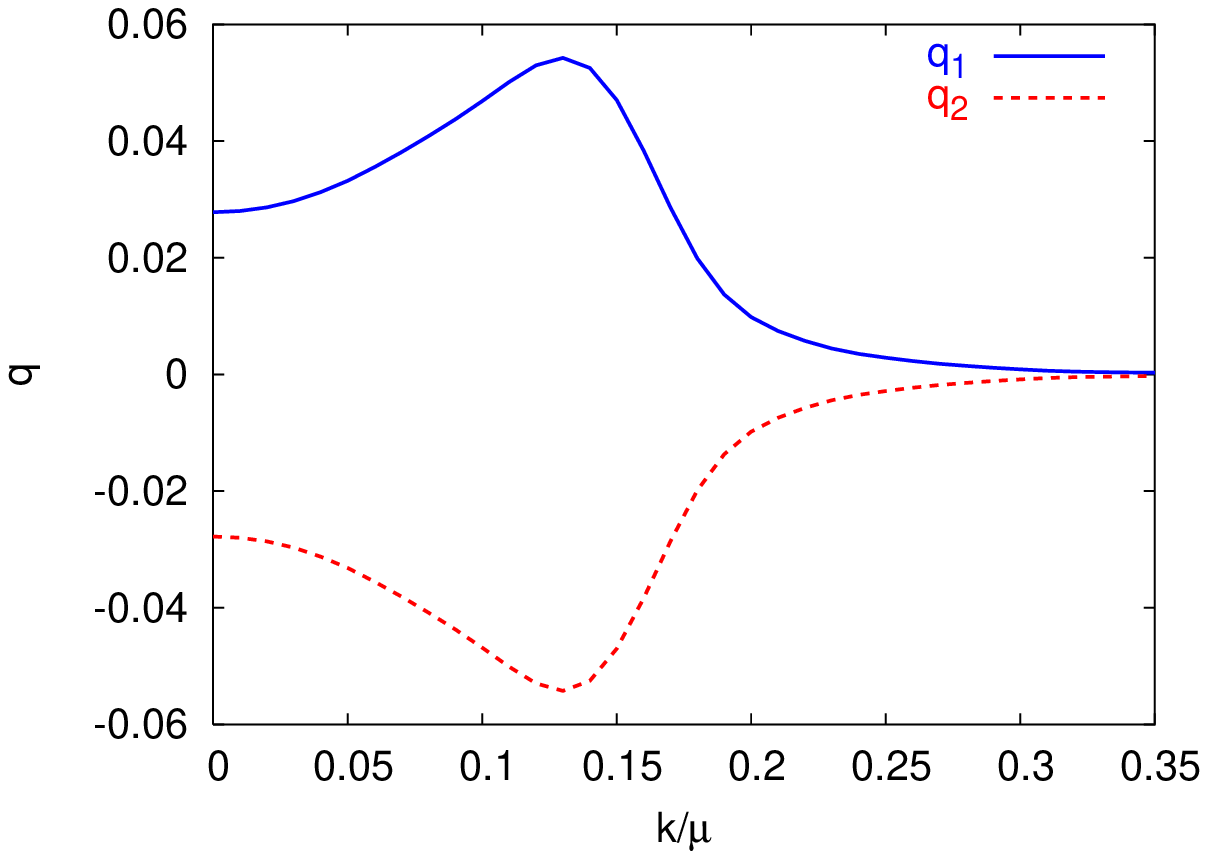, height=3.5in}
\end{center}
\vskip -0.35in
\caption{%
\small
The produced charges for the multiplet $({\bf 3},{\bf 2},\frac 16)$.
\label{figure:3216}
}
\end{figure}
\begin{figure}[htbp]
\begin{center}
\epsfig{file=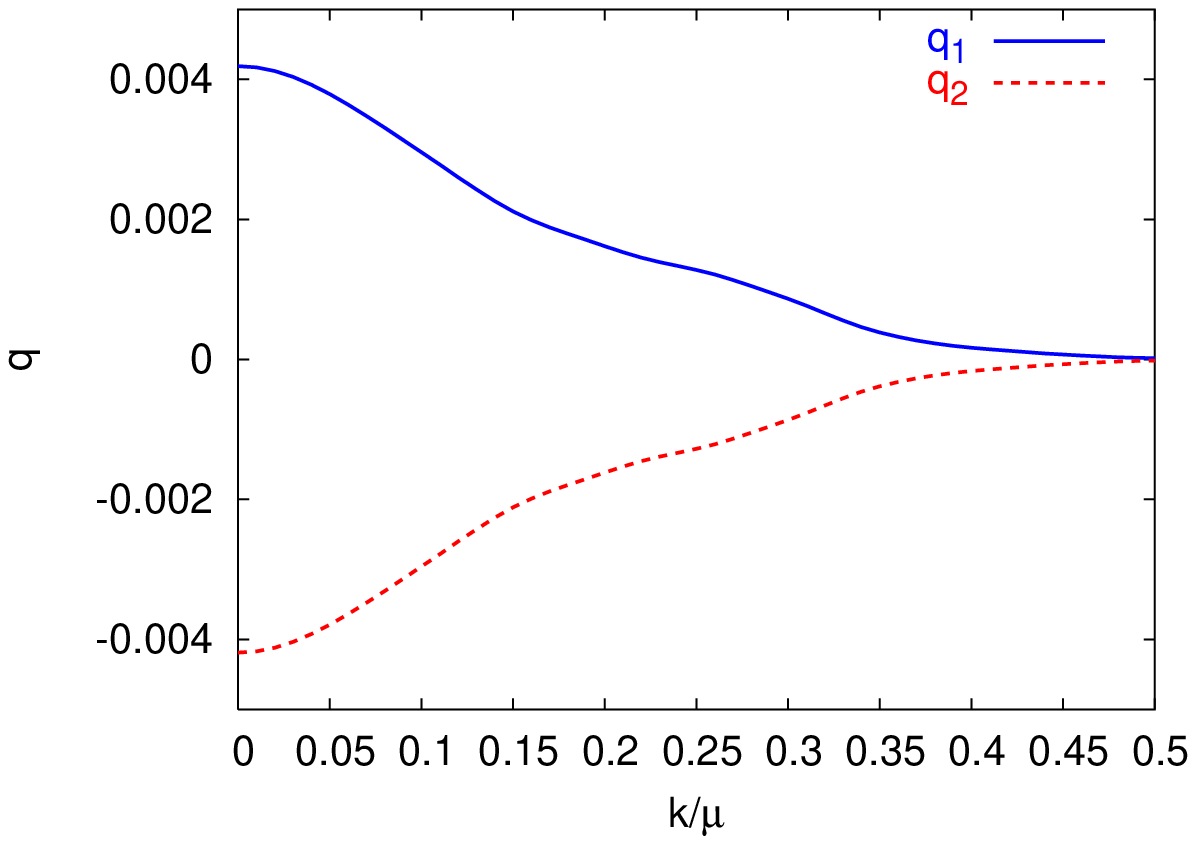, height=3.5in}
\end{center}
\vskip -0.35in
\caption{%
\small
The produced charges for the multiplet $({\bf\bar3},{\bf 1},-\frac 23)$.
\label{figure:3b1-23}
}
\end{figure}

The charge numbers which are plotted in
figures~\ref{figure:3216} and~\ref{figure:3b1-23}
refer to the mass matrix, which is diagonal after the phase transition is
completed. Hence, there is  mixing among $\psi_{\lambda_K}$,
$\psi_{A_K}$, $\psi_{C_K}$ and $\psi_{C^\prime_K}$. The mass matrix
is diagonalized {\it via} a biunitary
transformation. It turns out that the higgsino-gaugino mass matrix
in the supersymmetric vacuum has two heavy and two light eigenvalues.
We only display the charge production corresponding to
the light mass eigenvalues which we label by $q_1$ and $q_2$. There
is no substantial charge asymmetry stored within the heavy flavours,
$|q_3|, |q_4|\ll|q_1|, |q_2|$.
The apparent symmetry $\sum\limits_{i=1}^4 q_i=0$ results from the overall
${\rm U}(1)$ symmetry of the fermionic fields and is a useful check for
the numerical results.
We shall assume here that the resonant decay into 
gauge bosons and scalar Higgs particles is 
not important. This can be justified by noting that  
the necessary conditions for a resonant decay
into bosons are not met for our choice of parameters. 

The $q_i$ are charges stored within the mass eigenstates which are Dirac
fermions of the generic mixing form
\begin{equation}
\Psi_i=
\left(
\begin{array}{c}
{_i\alpha}^L_{\lambda_K}\lambda_K +
{_i\alpha}^L_{A_K}\psi_{A_K} +
{_i\alpha}^L_{C_K}\psi_{C_K} +
{_i\alpha}^L_{C^\prime_K}\psi_{C^\prime_K}
\\
{_i\beta}^R_{\lambda_{\bar{K}}}\bar\lambda_{\bar{K}} +
{_i\beta}^L_{A_{\bar{K}}}\bar\psi_{A_{\bar{K}}} +
{_i\beta}^L_{\bar{C}_{\bar{K}}}\bar\psi_{\bar{C}_{\bar{K}}} +
{_i\beta}^L_{\bar{C}^\prime_{\bar{K}}}\bar\psi_{\bar{C}^\prime_{\bar{K}}}
\end{array}
\right)\,,
\quad
i=1,2,3,4\,,
\end{equation}
where the coefficients $_i\alpha^L_X$ and $_i\beta^R_X$ are determined
by the biunitary transformation diagonalizing the higgsino-gaugino mass-matrix.
For the example we discuss here, the transformation matrices and the
coefficients are determined numerically.

Now, we discuss how the $q_i$ charges get transformed to $B\!-\!L$ charge
stored within fermionic matter. Decay into SUSY Standard Model particles
can take place through the gaugino components $\lambda_K$, $\lambda_{\bar{K}}$
and through the Higgsinos $\psi_{C_K}$, $\psi_{\bar{C}_{\bar{K}}}$.
The relevant operator for gaugino decay from the gauge supermultiplet
Lagrangean is
\begin{equation}
\label{decay:gaugino}
\sqrt 2 g \left\{\lambda FF + {\rm h.c.}\right\}
\end{equation}
while the higgsinos decay through the dimension \emph{five} couplings
\begin{eqnarray}
\label{decay:SO10:1st}
&&^i\gamma_1 \frac{\bar C F \bar C F}{M_S}\,,\\
\label{decay:SO10:2nd}
&&^i\gamma_2 \frac{C \Gamma^a F C \Gamma^a F}{M_S}\,,\\
\label{decay:SO10:3rd}
&&^i\gamma_3 \frac{C \Gamma^a \Gamma^b F C \Gamma^a \Gamma^b F}{M_S}\,,
\end{eqnarray}
which are
added to the superpotential~(\ref{superpotential:BarrRaby}), and
where $F$ are the Standard Model matter fields and
the right-handed neutrino, contained in $\mathbf{16}$, and $\Gamma$ denotes
the operators defined in~(\ref{Gamma:1-5}) and~(\ref{Gamma:6-10}).
The index $i=1,2,3$ denotes the matter generation.

While the coupling~(\ref{decay:gaugino}) is universal for all three generations
of Standard Model matter, $^i\gamma_1$, $^i\gamma_2$ and $^i\gamma_3$ may be
different for the three generations. We assume that
${ ^3\gamma_{2,3}}\gg { ^2\gamma_{2,3}}\gg { ^1\gamma_{2,3}}$, such that only 
the $ ^3\gamma_{2,3}$
are of relevance for the decays. In contrast, we require that the
decays through $^2\gamma_1$ and $^3\gamma_1$ are not possible since
the corresponding right-handed neutrinos are heavier than the decaying
particle, such that only $^1\gamma_1$ is relevant.
We now argue of which order these couplings should be for realistic scenarios.

For nonthermal leptogenesis~\cite{LazaridesShafi:1991}, one usually assumes
that two of the three Majorana neutrinos from the different generations
of matter fermions are heavier than half of the mass
\begin{equation}
m_I=\sqrt{2}\kappa\mu
\end{equation}
of the inflaton fields, which are the $\nu^c$-like components of
$C$ and $\bar C$ and the singlet $S$, such that their decay
into right handed neutrinos is kinematically forbidden. 
Through the coupling~(\ref{decay:SO10:1st}) the right handed neutrionos
acquire Majorana masses
\begin{equation}
^im_{\nu^c}= {^i\gamma}_1\frac{\langle C \rangle^2}{M_S}\,,
\end{equation}
such that the requirement $^{2,3}m_{\nu^c} > m_I/2$ reads
\begin{equation}
\label{couplings:heavy}
^{2,3}\gamma_1>\frac{1}{\sqrt 2}\kappa \frac{M_S}{\mu}\,,
\end{equation}
where we have used $\langle C \rangle=\mu$. It appears to be reasonable
that also the
couplings $^{2,3}\gamma_{2,3}$ are of the same order, as we shall assume.

For the scenario we discuss, the lightest right-handed neutrino is important
for the reheating process. The coupling~(\ref{decay:SO10:1st}) also allows
for the decay of the inflaton fields at the rate
\begin{equation}
\Gamma_\nu=\frac{1}{8\pi} m_I 
\left(
\frac{ ^1\gamma_1\langle C \rangle}{M_S}
\right)^2\,.
\end{equation}
The Universe becomes
radiation dominated and entropy production stops, when $\Gamma_\nu=H$,
where $H$ denotes the Hubble expansion rate. The reheat temperature at this
time is
\begin{equation}
T_R=0.55 g_*^{-\frac 14}
\sqrt{\Gamma_\nu m_{\rm Pl}}\,,
\label{T-reh}
\end{equation}
and we take the estimate $g_*=220$, the number of relativistic degrees
of freedom after reheating. The value for the Planck mass is
$m_{\rm Pl}=1.2\times 10^{19}{\rm GeV}$. The mass of the lightest
right-handed neutrino is therefore proportional to the reheat temperature:
\begin{equation}
\label{lightestnu}
^1m_{\nu^c}=7.7 \times \frac{g_*^\frac{1}{4}}{\sqrt \kappa}
\frac{\mu}{m_{\rm Pl}} T_R\,.
\end{equation}

Taking for the highest reheat temperature
allowed by the
gravitino bound $T_R=10^{11}{\rm GeV}$ and for the parameters we use, we
find $^1m_{\nu^c}< 5\times 10^9 {\rm GeV}$ while
$^{2,3}m_{\nu^c}> 2\times 10^{14} {\rm GeV}$.
Therefore, a fortuitous hierarchy of five orders of magnitude for the
Majorana masses is required. This is usually assumed for scenarios of
nonthermal leptogenesis.

For the coherent baryogenesis model we consider here however,
we can also allow for all Majorana masses
to be larger than $m_I/2$. Under these circumstances, $S$ and the
$\nu^c$-like components of $C$ and $\bar{C}$
cannot decay through the term~(\ref{decay:SO10:1st}) into two right-handed
neutrinos but decay to three particles \emph{via} the
operators~(\ref{decay:SO10:2nd}) and~(\ref{decay:SO10:3rd}). Since these
processes involve no Majorana particles, leptogenesis is absent for this
scenario.

We also have to deal with the fact, that for $\alpha_K=0$, namely
for $K=({\bf 1},{\bf 1},1)$, $\psi_{C_K}$ and $\psi_{C^\prime_K}$
do not mix, {\it cf.} the mass matrix~(\ref{SO10:mass:matrix}). Therefore
we assume that also the fields $C^\prime$ and $\bar{C}^\prime$ may decay
through couplings of the above type, suppressed however by additional powers
of $\langle R \rangle/M_S$, where $R$ is some singlet with a VEV.

Note however, that for $K=({\bf 1},{\bf 1},1)$ the mass
matrix~(\ref{SO10:mass:matrix})
is block-diagonal, such that only the pairs
$\lambda_K$-$\psi_{C_K}$ and
$\psi_{A_K}$-$\psi_{C^\prime_K}$ are mixed. Only for the second pair, the
$CP$-violating parameters $\xi$ and $\zeta$ are relevant and
an asymmetry is generated, which vanishes however after the decay
into matter. 

The charges $q_i$ gets processed differently to $B\!-\!L$ when
$\Psi_i$ decays through its various components. Let us denote
the $B\!-\!L$ number resulting from the decay of a component $X$
of a $\Psi_i$ quantum by $T_X$. By the couplings~(\ref{decay:gaugino}),
the reactions
\begin{eqnarray}
\lambda_K \!\!\!&\longrightarrow&\!\!\!F_{\bar{K}}^*+{\nu^c}^*\\
\lambda_{\bar{K}} \!\!\!&\longrightarrow&\!\!\!F_K^*+{\nu^c}^*
\end{eqnarray}
are induced, where one of the particles on the right hand side is a scalar,
the other one a fermion. Due to its Majorana mass term coming from the
operator~(\ref{decay:SO10:1st}), the right handed neutrino $\nu^c$
is its own antiparticle and therefore carries effectively $B\!-\!L=0$
at tree-level.
The resulting $B\!-\!L$-charge is therefore the one stored within
$F_{\bar{K}}^*$ and $F_K^*$ and we find
\begin{eqnarray}
\label{decay:lambda3216}
T_{\lambda_K} \!\!\!&=&\!\!\! \frac 13\,,
\quad K=({\bf 3},{\bf 2},\frac 16) \,, \\
\label{decay:lambda3b1-23}
T_{\lambda_K} \!\!\!&=&\!\!\! -\frac 13\,,\quad
K=({\bf \bar 3},{\bf 1},-\frac{2}{3})\,,\\
\label{decay:lambda11-1}
T_{\lambda_K} \!\!\!&=&\!\!\!  1\,,\quad
K=({\bf 1},{\bf 1},1)\,,\\
\label{decay:lambda3b2-16}
T_{\lambda_{\bar{K}}} \!\!\!&=&\!\!\! -\frac 13\,,
\quad\bar K=({\bf \bar 3},{\bf 2},-\frac 16) \,, \\
\label{decay:lambda3123}
T_{\lambda_{\bar{K}}} \!\!\!&=&\!\!\! \frac 13\,,\quad
\bar K=({\bf 3},{\bf 1},\frac{2}{3})\,,\\
\label{decay:lambda111}
T_{\lambda_{\bar{K}}} \!\!\!&=&\!\!\!  -1 \,,\quad
\bar K=({\bf 1},{\bf 1},-1)\,.
\end{eqnarray}

Similarly, the coupling~(\ref{decay:SO10:1st}) allows for the decay reaction
\begin{equation}
\psi_{{\bar C}_{\bar K}} \longrightarrow F_K^* + \nu^{c*}\,,
\end{equation}
Hence, the charges get transformed to
\begin{eqnarray}
\label{decay:16bar3216}
T_{\bar{C}_{\bar K}} \!\!\!&=&\!\!\! -\frac 13\,,
\quad\bar K=({\bf \bar 3},{\bf 2},-\frac 16) \,, \\
\label{decay:16bar3b1-23}
T_{\bar{C}_{\bar K}} \!\!\!&=&\!\!\! \frac 13 \,,\quad
\bar K=({\bf 3},{\bf 1},\frac{2}{3})\,,\\
\label{decay:16bar111}
T_{\bar{C}_{\bar K}} \!\!\!&=&\!\!\!  - 1 \,,\quad
\bar K=({\bf 1},{\bf 1},-1)\,.
\end{eqnarray}

We can calculate the term~({\ref{decay:SO10:2nd}) $\propto \gamma_2$ using
the techniques explained in the appendix~\ref{app:SO10}.
It is however easier
to note that ({\it cf.} Ref.~\cite{FukuyamaIlakovacKikuchiMeljanacOkada:2004})
\begin{equation}
({\bf 3},{\bf 2},\frac 16)\otimes({\bf \bar 3},{\bf 1},\frac 13)
\supset({\bf 1},{\bf 2},\frac 12)\subset {\bf 10}\,,
\end{equation}
as well as
\begin{equation}
({\bf 1},{\bf 1},0)\otimes({\bf 1},{\bf 2},-\frac 12)
=({\bf 1},{\bf 2},-\frac 12)\subset {\bf 10}\,.
\end{equation}
The components of $\psi_{C_K}$ for $K=({\bf 3},{\bf 2},\frac 16)$ therefore
decay to
\begin{eqnarray}
\psi_{C_{d}} \!\!\!&\longrightarrow&\!\!\! d^{c*} + e^*\,,\\
\psi_{C_{u}} \!\!\!&\longrightarrow&\!\!\! d^{c*} + \nu^* \,,
\end{eqnarray}
where $\psi_{{\bar C}_{d}}$ denotes the $d$-quark like higgsino,
$\psi_{{\bar C}_{u}}$ the $u$-quark like one. The charges hence get transformed
as 
\begin{equation}
\label{decay:16via10}
T_{C_K}=\frac 43 \,,\; \textnormal{for}\;
K=({\bf 3},{\bf 2},\frac 16) \,.
\end{equation}

The $u^c$-quark like higgsino with
$K=({\bf \bar 3},{\bf 1},-\frac 23)$ decays through the
$\gamma_3$-coupling~(\ref{decay:SO10:3rd}). We note
({\it cf.} Ref.~\cite{FukuyamaIlakovacKikuchiMeljanacOkada:2004})
\begin{eqnarray}
({\bf \bar 3},{\bf 1},-\frac 23)\otimes({\bf \bar 3},{\bf 1},\frac 13)
=({\bf 3},{\bf 1},-\frac 13)
\oplus ({\bf \bar 6},{\bf 1},-\frac 13)\subset {\bf 120}\,,\\
({\bf 1},{\bf 1},0)\otimes({\bf \bar 3},{\bf 1},\frac 13)
=({\bf \bar 3},{\bf 1},\frac 13)
\subset {\bf 120}\,,
\end{eqnarray}
and therefore have the reaction
\begin{equation}
\psi_{C_{u^c}} \longrightarrow d^{c*} + d^{c*}\,,
\end{equation}
and the charge conversion 
\begin{equation}
\label{decay:16via120}
T_{C_K}=\frac 23 \,,\;\textnormal{for}\;
K=({\bf \bar 3},{\bf 1},-\frac 23) \,.
\end{equation}

Finally, the $e^c$ like higgsino
$K=({\bf 1},{\bf 1},1)$ turns into matter {\it via} the
$\gamma_2$-coupling~(\ref{decay:SO10:2nd}), as can be seen by
\begin{eqnarray}
({\bf 1},{\bf 1},1)\otimes({\bf 1},{\bf 2},-\frac 12)
=({\bf 1},{\bf 2},\frac 12)\subset {\bf 10}\,,\\
({\bf 1},{\bf 1},0)\otimes({\bf 1},{\bf 2},-\frac 12)
=({\bf 1},{\bf 2},-\frac 12)
\subset {\bf 10}\,.
\end{eqnarray}
Consequently, the decay reaction is
\begin{equation}
\psi_{C^\prime_{e^c}} \longrightarrow e^{*} + \nu^{*}\,,
\end{equation}
and the resulting asymmetry 
\begin{equation}
\label{decay:16via10:2}
T_{C_K}=2 \,,\; \textnormal{for}\; K=({\bf 1},{\bf 1},1) \,.
\end{equation}

The procedure to obtain the produced $B\!-\!L$-density is as follows:
We first integrate the produced charge numbers $q_i$ over momentum space
in order to obtain charge densities $Q_i$.
From the biunitary diagonalization of $M$ in the supersymmetric minimum
the contributions of $\lambda_K$, $\lambda_{\bar K}$, $\psi_{C_K}$ and
$\psi_{{\bar C}_{\bar K}}$ to the $\Psi_i$
are determined, which gives the branching ratios
and therefore the respective contributions for
the decays of the Dirac fermions to Standard Model matter.
As a formula, this reads
\begin{equation}
n_K^0=\sum\limits_{i=1}^4 Q_i
\frac{
3| _i\alpha_{\lambda_K}^L|^2 2 g^2 T_{\lambda_K}
+| _i\alpha_{C_K}^L|^2 \left(\frac{ ^3\gamma_j \langle C \rangle}{M_S}\right)^2 T_{C_K}
-3| _i\beta_{\lambda_{\bar K}}^R|^2 2 g^2 T_{\lambda_{\bar K}}
-| _i\beta_{{\bar C}_{\bar K}}^R|^2 \left(\frac{ ^1\gamma_1 \langle C \rangle}{M_S}\right)^2 T_{{\bar C}_{\bar K}}
}
{
3| _i\alpha_{\lambda_K}^L|^2 2 g^2
+| _i\alpha_{C_K}^L|^2 \left(\frac{ ^3\gamma_j \langle C \rangle}{M_S}\right)^2
+3| _i\beta_{\lambda_{\bar K}}^R|^2 2 g^2
+| _i\beta_{{\bar C}_{\bar K}}^R|^2 \left(\frac{ ^1\gamma_1 \langle C \rangle}{M_S}\right)^2
}
,
\end{equation}
where the factors of three come from the presence of three generations of
matter and
\begin{equation}
j=\left\{
\begin{array}{c}
2\quad {\rm for} \quad K=({\bf 3},{\bf 2},\frac 16)\\
3\quad {\rm for} \quad K=({\bf \bar{3}},{\bf 1},-\frac 23)\\
2\quad {\rm for} \quad K=({\bf 1},{\bf 1},1)
\end{array}
\right.
\,.
\end{equation}

The total $(B-L)$ number density produced at the phase transition is taking
account of the
multiplicity of colour and flavour given by
\begin{equation}
n^0_{B-L}=6 n^0_{({\bf 3},{\bf 2},\frac 16)}
+3 n^0_{({\bf \bar 3},{\bf 1},-\frac 23)}
\,.
\end{equation}
A study of the parametric dependence of the produced asymmetry is beyond
the scope of this paper, which is to show that coherent baryogenesis is viable
with the gauge group ${\rm SO}(10)$, and shall be discussed elsewhwere.
Therefore, we
content ourselves with presenting just one typical numerical example here.
The parameters yet to be specified are the $^3\gamma_{2,3}$ and
$^1\gamma_1$. We can set effectively $^1\gamma_1=0$ because it is either
very small due to the restrictions given by the reheat 
temperature~(\ref{T-reh}) and
the gravitino bound or, in the case of $^1m_{\nu^c}>m_I/2$ decays through
the coupling~(\ref{decay:SO10:1st}) do not take place. In accordance with
the relation~(\ref{couplings:heavy}) we choose
$^3\gamma_2= ^3\gamma_1 = 0.05 M_S/\mu$. Then, we find
\begin{equation}
n^0_{B-L}=2.5 \times 10^{-7} \mu^3\,.
\end{equation}

In order to estimate the baryon to entropy ratio, we express the entropy
density $s$ through the reheat temperature $T_R$ as
\begin{equation}
s=2\pi^2g_* T_R^3/45\,,
\end{equation}
and the Hubble expansion rate is given by
\begin{equation}
H=1.66\sqrt{g_*}\frac{T_R^2}{m_{\rm Pl}}\,,
\end{equation}
where $m_{\rm Pl}=1.22 \times 10^{19}{\rm GeV}$ is the Planck mass.

During the epoch of coherent oscillations, that is between the end of inflation
and the onset of radiation era,
the Universe is matter dominated and expands by a
factor
\begin{equation}
\frac{a}{a_0}=\left(\frac {H_0}{H}\right)^\frac 23\,,
\end{equation}
where $H_0$ is the expansion rate at the end of inflation, given by
\begin{equation}
H_0=\sqrt{\frac{8\pi}{3}\frac{V}{{m_{\rm Pl}}^2}}\,.
\end{equation}

Putting everything together, we find
\begin{equation}
\label{temperature:reheating}
\frac{n_B}{s}
\approx \frac 13 \frac{n^0_{B-L}}{s}\left(\frac{a_0}{a}\right)^3
\approx\frac 14 \frac{n^0_{B-L}}{V_0} T_R
\,,
\end{equation}
where we have taken account of a division by three
for sphaleron transitions promoting $(B\!-\!L)$ to $B$ asymmetry.

The value for the vacuum energy at the end of inflation is
$V_0=\kappa^2\mu^4$, and by Eqn.~(\ref{temperature:reheating}),
we find
\begin{equation}
\label{BAU:numeric}
\frac{n_B}{s}=
1.0 \times 10^{-10}\,,
\end{equation}
where we have chosen $T_R = 2\times 10^{10} {\rm GeV}$. Hence, it appears
that in order to get a BAU in accordance with observation, there has to be
a reheat temperature of order of the upper bound allowed by the requirement
that gravitinos shall not be overproduced. However, our estimate
of entropy production is rather crude. It is conceivable that initially
the decay of $C$, $\bar C$ and $S$ is enhanced due to tachyonic and
also fermionic preheating. This could lead to an initial radiation-like
equation of state~\cite{PodolskyFelderKofmanPeloso2005} or
shorten the matter-dominated era
and would therefore lead to less dilution of the initial asymmetry.

For the case $^1m_{\nu_C}<m_I/2$,
it is of interest to compare the result~(\ref{BAU:numeric}) with the baryon
asymmetry resulting from nonthermal leptogenesis, which is given
by~{\cite{SenoguzShafi:2003}
\begin{equation}
\frac{n_B}{s}=0.5 \frac{\epsilon_1}{m_I}T_R\,,
\end{equation}
where
\begin{equation}
\epsilon_1=2\times 10^{-10} \left(\frac{^1m_{\nu_C}}{10^6 {\rm GeV}}\right)
\left(\frac{m_{\nu_3}}{50 {\rm meV}}\right)
\end{equation}
is the maximal $CP$-violation which may arise from the
decay of the lightest right-handed neutrino~\cite{Hamaguchi:2002},
and $m_{\nu_3}$ denotes the heaviest mass eigenvalue of the light neutrino
mass matrix. 
When we assume $m_{\nu_3}=50{\rm meV}$ and
use the same parameters as for the coherent baryogenesis
example and Eqn.~(\ref{lightestnu}) for $^1m_{\nu_C}$, we find
$n_B/s=6 \times 10^{-12}$. Therefore our example corresponds to a point in
parameter space where coherent baryogenesis dominates over leptogenesis.
However, we expect that also the opposite case may occur for a different
set of parameters.

\section{Conclusions}

In this paper,
we show that during the phase transition terminating
SUSY ${\rm SO}(10)$ hybrid inflation,
a charge asymmetry within the Higgsino sector may be produced 
through the mechanism of coherent baryogenesis and subsequently 
turned into baryons. $CP$-violation is provided by the couplings of the
spinorial to the adjoint representations and occurs at tree-level.
This is very different from leptogenesis, a one loop effect,
where $CP$-violation is sourced
by the matrix of Yukawa couplings between the neutrinos and the Standard Model
Higgs field.
Since the spinor-adjoint couplings are an
indispensable part of the Barr-Raby model, coherent baryogenesis naturally
occurs at the end of SUSY-${\rm SO}(10)$ hybrid inflation.
Together with similar results which we found for the Pati-Salam
group~\cite{GarbrechtProkopecSchmidt:2004}, this indicates that effects from
fermionic preheating are generically of importance for the generation of the
BAU in hybrid-inflationary scenarios.

Coherent baryogenesis however has been neglected in the
standard picture of baryogenesis in SUSY-GUT hybrid inflation
so far, which is as follows~\cite{LazaridesShafi:1991}:
The inflaton decays into right-handed neutrinos which then decay out-of
equilibrium into Standard Model matter,
leaving behind a $B\!-\!L$ asymmetry \emph{via} the leptogenesis
mechanism~\cite{FukugitaYanagida:1986}. Since the decaying Majorana neutrinos
are not produced by the thermal background, this scenario is often
called nonthermal leptogenesis.
However, it imposes strong constraints on the hierarchy of
the masses of right-handed neutrinos, as we discuss in section~\ref{sec:sim}.
We also emphasize that coherent baryogenesis relaxes this constraint and
allows
for all Majorana masses to be larger than the inflaton mass, since the
mechanism does not rely on leptogenesis and the decay of Majorana particles.
In conclusion, the relations of the parameters of hybrid inflationary models
to the BAU as suggested \emph{e.g.} in
Refs.~\cite{LazaridesShafi:1991,JeannerotKhalilLazaridesShafi:2000,KyaeShafi:2005} by considering leptogenesis, should be altered, since coherent baryogensis
turns out to be an additional source of the BAU, which may dominate in some
regions of parameter space.

We emphasize that nonthermal leptogenesis and coherent baryogenesis should not
be confused with the often discussed thermal
leptogenesis mechanism~\cite{FukugitaYanagida:1986,BuchmullerPlumacher:2000,BuchmullerPecceiYanagida:2005,Pilaftsis:1997,PilaftsisUnderwood:2003},
an appealing feature of which is that the BAU is generated
from a Universe which is -- within horizon scale -- in thermal equilibrium.
Leaving aside primordial density fluctuations, all cosmological observables
including the BAU would then be predictable from an effective theory valid
up to the Majorana mass scale of the neutrinos.

Grand Unified Theories open up many possible paths for the generation of the
BAU and it is yet not known where the actual asymmetry originates from. The
various mechanisms allow to establish relations to the paramters of the
underlying models. While leptogenesis renders constraints on the
neutrino sector, coherent baryogenesis is of interest since it is a scenario
of GUT-baryogenesis and thereby related to the dynamics of symmetry
breaking.

\section*{Acknowledgements}
We would like to thank Qaisar Shafi for interesting discussions
and Zurab Tavartkiladze for useful comments on the
manuscript.

\begin{appendix}

\section{${\rm SO}(10)$\label{app:SO10}}

Besides by tensors, orthogonal groups may also be represented by spinors,
which satisfy a Clifford algebra. In order to construct group-transformation
invariants, both types of representations need to be linked together {\it via}
Dirac gamma matrices.
For the familiar case of the Poincar\'e group $\textnormal{SO}(3,1)$,
it is often convenient
to use a specific representation for these matrices. In contrast, one
better circumvents the tedious task of explicitly constructing
ten $32 \times 32$ gamma matrices for $\textnormal{SO}(10)$.
Mohapatra and Sakita~\cite{MohapatraSakita:1979}
have therefore devised a very useful technique for performing calculations
involving spinors and tensors,
employing just abstract commutation and anticommutation relations.

On the other hand, when it comes to symmetry breaking, one has to choose
a certain convention, that is a certain basis,
how the particles of the Standard Model are assigned
to the representation $\mathbf{16}$ of $\textnormal{SO}(10)$. This assignment
fixes in turn the definition of the charge operators and hence
the quantum numbers of certain entries in vectors and tensors of
$\textnormal{SO}(10)$.

While the paper by Mohapatra and Sakita~\cite{MohapatraSakita:1979}
does not provide much details of tensor representations and symmetry breaking,
such a discussion can
be found in the comprehensive work by
Fukuyama {\it et~al.}~\cite{FukuyamaIlakovacKikuchiMeljanacOkada:2004}, where
in turn spinors are neglected. The coupling of spinors to tensors is
explained for ${\rm SO}(10)$ by Nath and Syed~\cite{NathSyed:2001}.
In the paper by Barr and Raby~\cite{BarrRaby:1997}, which contains the model we
consider here, a  basis where tensors nicely
decompose into blocks of $\textnormal{SU}(5)$-representations is chosen.
Unfortunately, the choice of basis and normalizations is not explicitly
given, but has to be inferred by the reader.

In the following, we give some detailed account of the construction of
$\textnormal{SO}(10)$-singlets, following the conventions of Barr and
Raby. Explicit expressions for the charge operators acting on spinors and
tensors as well as for the
accommodation of the Standard Model particles and the right-handed neutrino
in the representation $\mathbf{16}$ are given, which shall ensure an easier
and faster comprehensibility of the
Barr and Raby analysis as well as of our calculations.

\subsection*{Charge Assignments}

We denote by $Q$ the electric charge, by $Y$ the weak hypercharge and
by $I^3_L$ the weak isospin. The charges which are not gauge symmetries
of the Standard Model are baryon minus lepton number
$B-L$
as well as the $\textnormal{SU}(2)_R$-isospin $I^3_R$ and the less
known charge $X$.
There are linear dependencies among these charges, which are given by

\begin{eqnarray}
Q\!\!\!&=&\!\!\! I^3_L+Y\label{ChargeRel}\,,\\
B-L\!\!\!&=&\!\!\! 2(Y-I^3_R)\nonumber\,,\\
B-L\!\!\!&=&\!\!\! \frac 15 (4Y-X)\nonumber\,.
\end{eqnarray}
Note that, when comparing to the conventions by
Fukuyama \emph{et.~al.}~\cite{FukuyamaIlakovacKikuchiMeljanacOkada:2004},
we have twice as large values
for $(B\!-\!L)$, such that for a single lepton, we have $(B\!-\!L)=-1$.

In table~\ref{SMParticles}, we give the charge numbers of the Standard Model
particles and of the right-handed neutrino.

\begin{table}
\caption{Quantum numbers of matter\label{SMParticles}}
\begin{center}
\begin{tabular}{|c||c|c|c|c|c|c|}
\hline
& $Q$ & $I^3_L$ & $I^3_R$ & $Y$ & $B-L$ & $X$\\
\hline\vspace{-0.22cm}&&&&&&\vspace{-0.3cm}\\\hline
\vspace{-0.5cm}
&&&&&&\\
$Q=\left(
\begin{array}{c}
u\\
d
\end{array}
\right)$
&
$\begin{array}{c}
2/3\\
-1/3
\end{array}$
&
$\begin{array}{c}
1/2\\
-1/2
\end{array}$
&
$\begin{array}{c}
0\\
0
\end{array}$
&
$\begin{array}{c}
1/6\\
1/6
\end{array}$
&
$\begin{array}{c}
1/3\\
1/3
\end{array}$
&
$\begin{array}{c}
-1\\
-1
\end{array}$
\\
\vspace{-0.5cm}
&&&&&&\\
\hline
$\begin{array}{c}
u^c\\
d^c
\end{array}
$
&
$\begin{array}{c}
-2/3\\
1/3
\end{array}$
&
$\begin{array}{c}
0\\
0
\end{array}$
&
$\begin{array}{c}
-1/2\\
1/2
\end{array}$
&
$\begin{array}{c}
-2/3\\
1/3
\end{array}$
&
$\begin{array}{c}
-1/3\\
-1/3
\end{array}$
&
$\begin{array}{c}
-1\\
3
\end{array}$
\\
\hline
\vspace{-0.5cm}
&&&&&&\\
$L=\left(
\begin{array}{c}
\nu\\
e
\end{array}
\right)$
&
$\begin{array}{c}
0\\
-1
\end{array}$
&
$\begin{array}{c}
1/2\\
-1/2
\end{array}$
&
$\begin{array}{c}
0\\
0
\end{array}$
&
$\begin{array}{c}
-1/2\\
-1/2
\end{array}$
&
$\begin{array}{c}
-1\\
-1
\end{array}$
&
$\begin{array}{c}
3\\
3
\end{array}$
\\
\vspace{-0.5cm}
&&&&&&\\
\hline
$\begin{array}{c}
\nu^c\\
e^c
\end{array}
$
&
$\begin{array}{c}
0\\
1
\end{array}$
&
$\begin{array}{c}
0\\
0
\end{array}$
&
$\begin{array}{c}
-1/2\\
1/2
\end{array}$
&
$\begin{array}{c}
0\\
1
\end{array}$
&
$\begin{array}{c}
1\\
1
\end{array}$
&
$\begin{array}{c}
-5\\
-1
\end{array}$
\\
\hline
\end{tabular}
\end{center}
\end{table}

\subsection*{$\textnormal{SO}(2N)$ in an $\textnormal{SU}(N)$ Basis
\label{SO(2N)inSU(N)}}

This section contains a review of the paper by
Mohapatra and Sakita~\cite{MohapatraSakita:1979},
but adopts the basis conventions of Barr and Raby~\cite{BarrRaby:1997}.

Let us introduce $N$ operators $\chi_i \,\, (i=1,...,N)$,
acting on an antisymmetric Fock space, which obey the following
anticommutation relations:
\begin{eqnarray}
\{\chi_i,\chi_j^\dagger\}\!\!\!&=&\!\!\!\delta_{ij}\label{chi_acomm:1}\,,\\
\{\chi_i,\chi_j\}\!\!\!&=&\!\!\!0\label{chi_acomm:2}\,.
\end{eqnarray}
The operators defined as
\begin{equation}
T^{i}_{\phantom{i}j}=\chi^\dagger_i \chi_j
\end{equation}
satisfy the $\textnormal{SU}(N)$ algebra:
\begin{equation}
[T^{i}_{\phantom{i}j},T^{k}_{\phantom{k}l}]=
\delta^k_{\phantom{k}j}T^{i}_{\phantom{i}l}-
\delta^i_{\phantom{i}l}T^{k}_{\phantom{k}j}\,.
\end{equation}

We now introduce the $2N$ operators
\begin{eqnarray}
\label{Gamma:1-5}
\Gamma_j\!\!\!&=&\!\!\!-{\rm i} (\chi_j-\chi_j^\dagger)\,,
\quad j=1,...,N\,,\\
\label{Gamma:6-10}
\Gamma_{N+j}\!\!\!&=&\!\!\!\chi_j+\chi_j^\dagger\,,
\end{eqnarray}
which obey by Eqns.~(\ref{chi_acomm:1},~\ref{chi_acomm:2})
the Clifford algebra
\begin{equation}
\{\Gamma_i,\Gamma_j\}=2\delta_{ij}\,,\quad i,j=1,...,2N\,,
\end{equation}
and hence, the algebra of generators of $\textnormal{SO}(2N)$ is given by
\begin{equation}
\label{SO10:generators}
\Sigma_{ij}=\frac{1}{2{\rm i}}[\Gamma_i,\Gamma_j]\,.
\end{equation}
Since the dimension of the spinor representation of $\textnormal{SO}(2N)$
is $2^N$, a concrete representation could be constructed
for $\textnormal{SO}(10)$ in terms of
$32 \times 32$-matrices, which however shall not be done here.

The spinor states can be constructed by letting the $N$ creation operators
$\chi_i^\dagger$ act on the ``vacuum'' $|0\rangle$, such that the spinor
representation is $2^N$-dimensional, as it should.

It is well known, that the spinor representation of
$\textnormal{SO}(2N)$ is reducible. We therefore define
\begin{equation}
\Gamma_0={\rm i}^N\prod_{i=1}^{2N} \Gamma_i=\prod_{j=1}^{N}(1-2n_j)\,,
\end{equation}
where we have introduced the number operators
\begin{equation}
n_j=\chi_j^\dagger \chi_j\,.
\end{equation}
The chiral projectors $\frac 12 (1 \pm \Gamma_0)$ give therefore rise
to the two irreducible $2^{N-1}$-dimensional representations containing only
even~(case ``$+$'') or only odd~(case ``$-$'') numbers of creation operators.

Now let $\Psi$ be an $\textnormal{SO}(2N)$ spinor state. We are interested
in calculating products of the form
\begin{equation}
\Psi^T B  \Gamma_{i_1} ... \Gamma_{i_M} \Psi\,,
\end{equation}
involving a certain number of $\Gamma$ matrices. The matrix $B$ is necessary
since $\Psi^T$ does not transform as a conjugate spinor when acted upon with
an infinitesimal $\textnormal{SO}(10)$-transformation $\epsilon_{ij}$:
\begin{eqnarray}
\delta \Psi\!\!\!&=&\!\!\! {\rm i}\epsilon_{ij}\Sigma_{ij}\Psi\,,\\
\delta \Psi^\dagger\!\!\!&=&\!\!\! -\rm{i}\epsilon_{ij}\Psi^\dagger\Sigma_{ij}\nonumber\,,\\
\delta \Psi^T\!\!\!&=&\!\!\! {\rm i}\Psi^T\epsilon_{ij}\Sigma_{ij}^T
\,.
\nonumber
\end{eqnarray}
We require from $B$ the property
\begin{equation}
B^{-1}\Sigma_{ij}^T B=-\Sigma_{ij}\label{Sigma:B:Trafo}\,,
\end{equation}
such that
\begin{equation}
\delta(\Psi^T B)={\rm i}\epsilon_{ij} \Psi^T B B^{-1}\Sigma_{ij}^T B=
-{\rm i}\epsilon_{ij} (\Psi^T B) \Sigma_{ij}\,,
\end{equation}
{\it i.e.} $\Psi^T B$ transforms as a conjugate spinor. The
condition~(\ref{Sigma:B:Trafo}) can be met if
\begin{equation}
B^{-1}\Gamma_i^TB=\pm \Gamma_i\,.
\end{equation}
By choosing the minus-sign in the latter equation, we find
\begin{equation}
B=\prod_{i=1}^N \Gamma_i\,,
\end{equation}
because for $i=1,...,N$ the $\Gamma_i$ are represented by antisymmetric matrices,
while for $i=N+1,...,2N$ by symmetric ones.

For $N=5$, we can arrange the Standard Model particles in the
spin-$\mathbf{16}$ representation, which is projected out of the
$32$-dimensional
spinor $\Psi$ by $\frac 12 (1-\Gamma_0)\Psi$. Defining
\begin{eqnarray}
\label{SM:Particles}
u_i\!\!\!&=&\!\!\!\frac 12 \varepsilon^{ikl45} 
  \chi_k^\dagger \chi_l^\dagger \chi_5^\dagger |0\rangle \,, \\
d_i\!\!\!&=&\!\!\!\frac 12 \varepsilon^{ikl45} 
  \chi_k^\dagger \chi_l^\dagger \chi_4^\dagger |0\rangle \,, \nonumber\\
u_i^c\!\!\!&=&\!\!\!\chi_i^\dagger \chi_4^\dagger \chi_5^\dagger |0\rangle \,, \nonumber\\
d_i^c\!\!\!&=&\!\!\!\chi_i^\dagger|0\rangle \,, \nonumber\\
\nu\!\!\!&=&\!\!\!\chi_5^\dagger|0\rangle \,, \nonumber\\
e\!\!\!&=&\!\!\!\chi_4^\dagger|0\rangle \,, \nonumber\\
\nu^c\!\!\!&=&\!\!\!\chi_1^\dagger\chi_2^\dagger\chi_3^\dagger\chi_4^\dagger\chi_5^\dagger
             |0\rangle \nonumber \,,\\
e^c\!\!\!&=&\!\!\!\chi_1^\dagger\chi_2^\dagger\chi_3^\dagger
             |0\rangle \nonumber \,,
\end{eqnarray}
where $i,k,l=1,2,3$. ~\emph{Cf.} also Ref~\cite{WilczekZee:1981}, where the
doublet and triplet blocks are interchanged.

The next task is to construct the charge operators. For example, the ladder
operators associated with the left isospin take $u \leftrightarrow d$ and
$\nu \leftrightarrow e$. They are therefore given by
\begin{eqnarray}
I_L^+ \!\!\!&=&\!\!\! \chi_5^\dagger \chi_4\,,\\
I_L^- \!\!\!&=&\!\!\! \chi_4^\dagger \chi_5 \,.\nonumber
\end{eqnarray}
The weak isospin operator is hence
\begin{eqnarray}
I_L^3=\frac 12 [I_L^+,I_L^-]=\frac 12 (n_5-n_4).
\end{eqnarray}

By comparison with the charge numbers in table~\ref{SMParticles}, we can
identify
\begin{equation}
Y=\frac 13 \sum_{i=1}^3 n_i- \frac 12 \sum_{j=4}^5 n_j=
\frac{1}{12{\rm i}}\left(
[\Gamma_1,\Gamma_6]+[\Gamma_2,\Gamma_7]+[\Gamma_3,\Gamma_8]
\right)
-\frac{1}{8{\rm i}}\left(
[\Gamma_4,\Gamma_9]+[\Gamma_5,\Gamma_{10}]
\right)\,,
\end{equation}
where we have used
\begin{equation}
[\Gamma_{5+j},\Gamma_j]=-4{\rm i}n_j+2{\rm i}\,.
\end{equation}
When identifying the indices of the $\Gamma$ operators with matrix
rows and columns as implied by Eqn.~(\ref{SO10:generators}),
we can explicitly write down the suitably normalized $Y$ in tensor
representation:
\begin{equation}
Y={\rm diag}\left(1/3,1/3,1/3,-1/2,-1/2\right)\otimes \sigma_2.
\end{equation}
Generally, we use as rule for conversion of the operator to the tensor
representation\footnote{This is of course strictly speaking no equality
but an assignment of an operator acting in Fock space to an operator
acting in tensor space.}
\begin{equation}
\label{Spinor:Tensor}
\left[\frac 14 [\Gamma_i,\Gamma_j]\right]_{ab}=
\delta_{ia}\delta_{jb}-\delta_{ja}\delta_{ib}\,,
\end{equation}
which reads for the special case of the charge operators
\begin{equation}
\frac {\rm i}{4} \left[\Gamma_{5+i},\Gamma_i\right]
=n_i-\frac 12 =
{\rm diag}\left(\delta_{1i},\delta_{2i},\delta_{3i},\delta_{4i},
\delta_{5i}\right)\otimes \sigma_2\,.
\end{equation}

Now, we easily find the other charge operators. Putting everything together,
we have in spinor and in tensor representation
\begin{eqnarray}
\label{charge:operators}
Q\!\!\!&=&\!\!\!\frac 13 \sum_{i=1}^3 n_i - n_4
={\rm diag}\left(1/3,1/3,1/3,-1,0\right)\otimes \sigma_2\,,\\
I^3_L\!\!\!&=&\!\!\!\frac 12 (n_5-n_4)
={\rm diag}\left(0,0,0,-1/2,1/2\right)\otimes \sigma_2
\nonumber\,,\\
I^3_R\!\!\!&=&\!\!\!\frac 12 (1-n_4-n_5)
={\rm diag}\left(0,0,0,-1/2,-1/2\right)\otimes \sigma_2
\nonumber\,,\\
B-L\!\!\!&=&\!\!\!\frac 23 \sum_{i=1}^3 n_i - 1
={\rm diag}\left(2/3,2/3,2/3,0,0\right)\otimes \sigma_2
\nonumber\,,\\
Y\!\!\!&=&\!\!\!\frac 13  \sum_{i=1}^3 n_i - \frac 12 \sum_{j=4}^5 n_j
={\rm diag}\left(1/3,1/3,1/3,-1/2,-1/2\right)\otimes \sigma_2
\nonumber\,,\\
X\!\!\!&=&\!\!\!-2  \sum_{i=1}^5 n_i +5
={\rm diag}\left(-2,-2,-2,-2,-2\right)\otimes \sigma_2
\nonumber
\,,
\end{eqnarray}
where we have used the normalization convention~(\ref{ChargeRel}).

The operator representation for the charge operators
$\mathcal{Q}$ is suitable
for finding the charge eigenvalues $q$ of the spinors through
$\mathcal{Q} \Psi= q \Psi$.

Tensors can be constructed from the fundamental 10-dimensional
vector $\Phi_{10}$ by taking antisymmetric products, such that a
rank $n$ tensor is of dimension
$10\cdot 9\cdot ...\cdot (10-n+1) /n!$.
Explicitly, for the vector and the rank two tensor, the charges
implied by the gauge-covariant derivatives are given by the eigenvalue
equations
\begin{eqnarray}
\mathcal{Q}_q \Phi_{10} \!\!\!&=&\!\!\! q \Phi_{10}\,,\\
{[}\mathcal{Q}_q, \Phi_{45}{]}\!\!\! &=&\!\!\! q \Phi_{45}\nonumber
\,,
\end{eqnarray}
where $\mathcal{Q}$ is acting here by matrix multiplication.

\subsection*{The Tensor Representations\label{Appendix:Tensors}}

In order to perform calculations such as
$\mathbf{16}.\mathbf{45}.\mathbf{\overline{16}}$, ${\rm tr} \mathbf{45}^4$
and $\mathbf{16}.\mathbf{10}.\mathbf{16}$,
we need to identify the Standard Model multiplets within 
$\mathbf{10}$ and $\mathbf{45}$,
just as we did for the $\mathbf{16}$ in~(\ref{SM:Particles}).
We first note, that under $\textnormal{SU}(5)$, the fundamental
representation of $\textnormal{SO}(10)$ decomposes as
$\mathbf{10}=\mathbf{5}\oplus\mathbf{\bar{5}}$. 
Let us denote an element of $\mathbf{5}$ in the representation
$\mathbf{10}$ of $\textnormal{SO}(10)$ by $\Phi_\mathbf{10}^\mathbf{5}$,
an element of $\mathbf{\bar{5}}$ by $\Phi_{\mathbf{10}}^\mathbf{\bar{5}}$.
Since they obey
\begin{equation}
X \Phi_\mathbf{10}^\mathbf{5} = 2 \Phi_\mathbf{10}^\mathbf{5}
\quad\textnormal{and}\quad
X \Phi_\mathbf{10}^\mathbf{\bar{5}} = -2 \Phi_\mathbf{10}^\mathbf{\bar{5}}
\,,
\end{equation}
they are of the form
\begin{equation}
\Phi_\mathbf{10}^\mathbf{5}=
\frac{1}{\sqrt{2}}
\left(
\begin{array}{c}
a_1\\\vdots\\a_5\\
-{\rm i}a_1\\\vdots\\-{\rm i}a_5
\end{array}
\right)
\quad\textnormal{and}\quad
\Phi_\mathbf{10}^\mathbf{\bar{5}}=
\frac{1}{\sqrt{2}}
\left(
\begin{array}{c}
\bar{a}_1\\\vdots\\\bar{a}_5\\
{\rm i}\bar{a}_1\\\vdots\\{\rm i}\bar{a}_5
\end{array}
\right)\,,
\end{equation}
with $\sum_{i=1}^5 |a_i|^2=1$ and $\sum_{i=1}^5 |\bar{a}_i|^2=1$.
To remove this inconvenient mixing of the upper and lower five-blocks,
we introduce the unitary transformation
\begin{equation}
\label{blocktrafo}
U_{\textnormal{\tiny BLOCK}}=\frac{1}{\sqrt{2}}
\left(
\begin{array}{cc}
\mathbbm{1}_5&{\rm i}\mathbbm{1}_5\\
\mathbbm{1}_5&-{\rm i}\mathbbm{1}_5 
\end{array}
\right)\,,
\quad
U_{\textnormal{\tiny BLOCK}}^{-1}=\frac{1}{\sqrt{2}}
\left(
\begin{array}{cc}
\mathbbm{1}_5&\mathbbm{1}_5\\
-{\rm i}\mathbbm{1}_5& {\rm i}\mathbbm{1}_5 
\end{array}
\right)\,,
\end{equation}
such that
\begin{equation}
U_{\textnormal{\tiny BLOCK}}\Phi_\mathbf{10}^\mathbf{5}=
\left(
\begin{array}{c}
a_1\\\vdots\\a_5\\
0\\\vdots\\0
\end{array}
\right)
\quad\textnormal{and}\quad
U_{\textnormal{\tiny BLOCK}}\Phi_\mathbf{10}^\mathbf{\bar 5}=
\left(
\begin{array}{c}
0\\\vdots\\0\\
\bar{a}_1\\\vdots\\\bar{a}_5
\end{array}
\right)\,.
\end{equation}

By this change of basis,
the charge operators become diagonal, for example
\begin{equation}
U_{\textnormal{\tiny BLOCK}} X U_{\textnormal{\tiny BLOCK}}^{-1}=
2\left(
\begin{array}{cc}
\mathbbm{1}_5&0\\
0&-\mathbbm{1}_5 
\end{array}
\right)\,.
\end{equation}

We can therefore immediately see how the entries of $\mathbf{45}$ transform
under $\textnormal{SU}(5)$, namely
\begin{equation}
U_{\textnormal{\tiny BLOCK}}
\Phi_{\mathbf{45}}
U_{\textnormal{\tiny BLOCK}}^{-1}
=
\left(
\begin{array}{c|c}
\mathbf{24}\oplus\mathbf{1}&\mathbf{10}\\
\hline
\mathbf{\overline{10}}&\mathbf{24}\oplus\mathbf{1}
\end{array}
\right)\,,
\end{equation}
where the single entries represent $5\times 5$-blocks and the blocks in the
upper left and the lower right are to be related to each other by the factor
of minus one. The $\textnormal{SU}(5)$-singlet~$\mathbf{1}$ has here
the form of the matrix $1/\sqrt5 \mathbbm{1}_5$.
The arrangement of the $G_{SM}$-multiplets contained in
$\mathbf{24}$ can be schematically written as
\vskip 0.1cm
\begin{equation}
\textnormal{
\begin{tabular}{|c||c|c|}
\hline
& $1\quad 2\quad 3$ & $4\quad 5$\\
\hline\vspace{-0.22cm}&&\vspace{-0.3cm}\\\hline
$\begin{array}{c}1\\2\\3\end{array}$
&$({\bf 8},{\bf 1},0)\oplus({\bf 1},{\bf 1},0)$
&$({\bf 3},{\bf 2},-\frac{5}{6})$\\
\hline
$\begin{array}{c}4\\5\end{array}$
&$({\bf \bar{3}},{\bf 2},\frac{5}{6})$&$({\bf 1},{\bf 3},0)$\\
\hline
\end{tabular}
}
\,;
\end{equation}
\vskip 0.1cm
\noindent
and finally $\mathbf{10}$ of ${\rm SU}(5)$ decomposes into
\vskip 0.1cm
\begin{equation}
\textnormal{
\begin{tabular}{|c||c|c|}
\hline
& $1\quad 2\quad 3$ & $4\quad 5$\\
\hline\vspace{-0.22cm}&&\vspace{-0.3cm}\\\hline
$\begin{array}{c}1\\2\\3\end{array}$
&$({\bf \bar{3}},{\bf 1},-\frac{2}{3})$
&$({\bf 3},{\bf 2},\frac{1}{6})$\\
\hline
$\begin{array}{c}4\\5\end{array}$
&$-({\bf 3},{\bf 2},\frac{1}{6})$&$({\bf 1},{\bf 1},1)$\\
\hline
\end{tabular}
}
\,,
\end{equation}
\vskip 0.1cm
\noindent
where the matrix is imposed to be antisymmetric, since it is identified
with the antisymmetric part of $\mathbf{5}\otimes\mathbf{5}$ of
$\textnormal{SU}(5)$.


\end{appendix}

\end{document}